\documentclass[aoas,nameyear,MSNbibl,dvips]{arximspdf}
\usepackage{graphicx}
\usepackage{breakurl}

\doi{10.1214/10-AOAS398E}
\referstodoi{10.1214/10-AOAS398}
\volume{5}
\issue{1}
\pubyear{2011}
\firstpage{83}
\lastpage{87}

\makeatletter

\def\@bmisc[#1]{%
  \get@battribute{unstr}%
  \common@pub@types%
  \let\bauthor\bbl@bauthor%
  \let\bhowpublished\@firstofone%
  \def\borganization##1{{\bauthor@style ##1}}%
}

\renewcommand{\epsilon}{\varepsilon}
\makeatother

\begin{document}
\begin{frontmatter}

\title{Spurious predictions with random time series: The~Lasso in the
context of paleoclimatic reconstructions.
Discussion of: A statistical analysis of multiple temperature proxies: Are
reconstructions of surface temperatures over the last 1000 years reliable?}
\runtitle{Discussion}
\pdftitle{Discussion on A statistical analysis of multiple temperature
proxies: Are reconstructions of surface temperatures over the last 1000
years reliable?
by B. B. McShane and A. J. Wyner}

\begin{aug}
\author{\fnms{Martin P.} \snm{Tingley}\corref{}\thanksref{a1}\ead[label=e1]{tingley@fas.harvard.edu}}

\runauthor{M. P. Tingley}

\affiliation{NCAR and Harvard University}

\address{Institute for Mathematics Applied to Geosciences\\
National Center for
Atmospheric Research\\
1850 Table Mesa Drive\\
Boulder, Colorado 80305\\
USA\\
\printead{e1}} %adresu isvedimo komanda gale!
\end{aug}
\thankstext{a1}{A more detailed version of this discussion is available at
\protect\href{http://people.fas.harvard.edu/\textasciitilde tingley/}{http://people.fas.harvard.edu/}
\protect\href{http://people.fas.harvard.edu/\textasciitilde tingley/}{\textasciitilde tingley/}.}

% HISTORY:
\received{\smonth{9} \syear{2010}}

% ABSTRACT

% KEYWORDS

\end{frontmatter}

Blakeley B. McShane and Abraham J. Wyner (hereafter, MW2011) find that,
under certain scenarios and using the LASSO to fit regression models,
randomly generated series are as predictive of past climate as the
commonly used proxies (MW2011, Figure 9). They conclude that ``the
proxies do not predict temperature significantly better than random
series generated independently of temperature,'' a~claim that has
already been reproduced in the popular press \citep
{WSJ2010McShaneQuote}. If this assertion is correct, then MW2011 have
undermined all efforts to reconstruct past climate, which are based on
the fundamental assumption that natural proxies are predictive of past
climate. I disagree with MW2011's conclusion and provide an alternative
explanation: the LASSO, as applied in MW2011, is simply not an
appropriate tool for reconstructing paleoclimate.

To shed light on the MW2011 results, I turn to an experiment with
surrogate data \citep{TingleySuppMcShane}. The ``target'' time series,
analogous to the Northern Hemisphere mean temperature time series in
MW2011, is the sum of a simple linear trend and an AR(1) process,
$y(t)= 0.25\cdot t+\epsilon(t), t=1, \ldots,149$. The AR(1) coefficient in
the $\epsilon$ process is $0.4$, and the variance of the innovations is
1. I then generate 1138 ``pseudo-proxy'' time series by adding white
noise to this target series. The signal to noise ratio (SNR) of these
pseudo-proxies, expressed as the ratio of the standard deviation of the
target time series to that of the additive white noise, will take on a
range of values $(4, 2, 1, 1/2, 1/4, 1/8)$. In order to compare the
performance of these pseudo-proxies to random series, I generate 1138
independent AR(1) time series, each of length 149; the common AR(1)
coefficient, $\alpha$, for these random series will take on a range of
values (0, 0.2, 0.4, 0.6, 0.8, 1.0). Two regression models are then fit
using 119 of the 149 observations.

The first model, referred to as ``composite regression,'' involves
averaging across all predictor series and then using this composite
series to predict the target via ordinary least squares regression. The
second model applies the LASSO to all predictor series, and is fit
using the algorithm described in \citeauthor{Friedman2007p1282} (\citeyear{Friedman2007p1282,Friedman2010p1281})
and the \texttt{glmnet} package for Matlab
(available at
\url{http://www-stat.stanford.edu/\textasciitilde tibs/glmnet-matlab/}).
The LASSO penalization parameter ($\lambda$ on page 13 of MW2011) is
set to be 0.05 times the smallest value of $\lambda$ for which all
coefficients are zero; the LASSO penalization is thus very small.

%f1 ###
\begin{figure}

\includegraphics{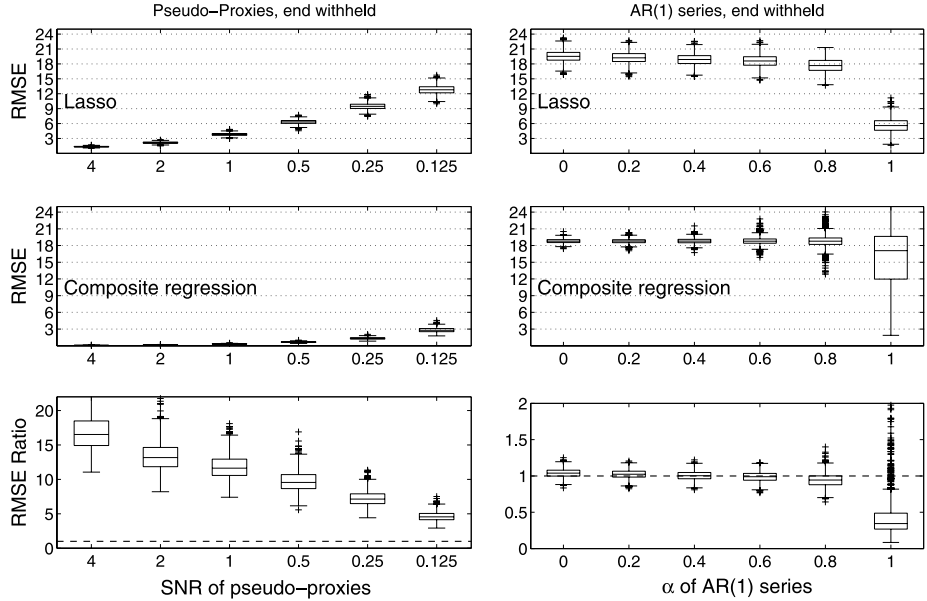}

\caption{Out-of-sample RMSE calculated using 30 values withheld from
the end of each surrogate data set. Left column: using pseudo-proxies
as predictors. Right column: using independent AR(1) series as
predictors. \textup{Top row:} regression using the LASSO. \textup{Middle row:} composite
regression. \textup{Bottom row:} the ratio of the LASSO RMSE value to the
composite regression RMSE.}
\label{End_wh_6subs}
\end{figure}

Box plots of the out-of-sample RMSE are shown in Figure~\ref
{End_wh_6subs} for 1000 experiments that calculate the RMSE using
observations withheld from the end of the data set; results are similar
when observations are withheld from the interior. Composite regression
results in lower RMSE than the LASSO for all values of the pseudo-proxy
SNR (Figure~\ref{End_wh_6subs}, left column). For an SNR of $1/4$, the
LASSO RMSE is about 7.5 times larger than the composite regression
RMSE. This is a clear indication that the LASSO is not making effective
use of the information contained in the pseudo-proxies.

\begin{figure}

\includegraphics{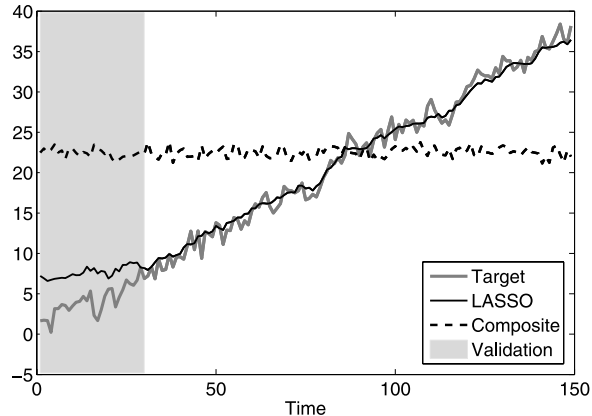}

\caption{Example fits from applying the LASSO to random walk predictors
and composite regression to white noise predictors. Shading indicates
the portion of the data set withheld for validation.}
\label{fig:AR_DEMO}
\end{figure}

Applying the LASSO to AR(1) series with sufficiently high $\alpha$
values results in lower out-of-sample RMSE values than applying the
LASSO to the noisier pseudo-proxies (compare the two top panels of
Figure~\ref{End_wh_6subs}). This is the result discussed in MW2011: the
LASSO gives better results when applied to highly structured random
time series than when applied to noisy predictors that do in fact
contain information about the target series. Note in addition that for
values of $\alpha\geq0.8$, the LASSO on the AR(1) series results in
lower RMSE than using composite regression on AR(1) series with $\alpha
=0$ (the limiting case of an SNR of zero for the pseudo-proxies). These
results can be explained by the structure of the surrogate data
experiment, which sets the target series to be linear in time, with
additive AR(1) noise. The LASSO applied to AR(1) series with $\alpha=1$
results in nonzero coefficients for only those predictor series that
display strong, linear trends over the calibration interval, and the
expected value of a predictor series during the validation interval is
then the last value in the calibration interval. In contrast, as the
SNR${} \rightarrow0$, composite regression on the pseudo-proxies
approaches (in expectation) the intercept model. These features are
illustrated in Figure~\ref{fig:AR_DEMO}.

%f2 ###

MW2011 point out that highly structured random series (large $\alpha$)
are well suited to interpolation, and to a lesser extent extrapolation,
on short time scales. As the variance of the white noise component of
the pseudo-proxies increases, these predictors become both less
informative of the target series, and less structured in time. At a
certain SNR, short-term interpolations or extrapolations based on
independent, but more temporally structured series, perform better.
This threshold SNR is a decreasing function of the length of the
extrapolation/interpolation interval. As the goal in a paleoclimate
context is extrapolation on long timescales, composite regression on
extraordinarily noisy proxies will outperform the LASSO applied to
random walks.

The LASSO gives inferior results in situations where each of a large
number of predictors is only weakly correlated with the target series,
but the mean across all predictors is highly correlated with that
target. It is well known that the LASSO is the posterior mode which
results from placing a common double exponential prior on the
regression coefficients [\citet{Park2008p2787}]. It is difficult to
imagine a scientifically defensible reason for specifying such a prior
in the paleoclimate context. A more scientifically reasonable approach
is to modify the LASSO prior to shrink the regression coefficients not
towards zero, but toward a common, data determined value. Such a prior
reflects the assumptions that (1) the regression coefficients are
likely to be similar to one another, and (2) all predictors are
informative of the target series. Within the paleoclimate context,
where the expectation is that each proxy is weakly correlated to the
northern hemisphere mean (for two reasons: proxies generally have a
weak correlation with local climate, which in turn is weakly correlated
with a hemispheric average) the LASSO as used by MW2011 is simply not
an appropriate tool. It throws away too much information.

More generally, MW2011 have perhaps missed a larger point. The presence
of a large number of correlated predictors is intrinsic to the
paleoclimate reconstruction problem and has a geophysical basis. MW2011
state that, ``it is unavoidable that some type of dimensionality
reduction is necessary, even if there is no principled way to achieve
this.'' This is simply not the case. A more scientifically sound
approach recognizes that the proxies are related to the local climate,
which in turn displays both spatial and temporal correlation. These
ideas can be encoded in hierarchical statistical models, which can
combine the specification of a parametric spatiotemporal covariance
form for the target climate process (e.g., surface temperature
anomalies) with reasonable forward models that describe the conditional
distribution of the proxy observations given the climate process. Such
approaches naturally account for the $p\gg n$ problem, and for the
strong correlations between the proxies. These models are derived from
the rich development of Bayesian statistics over the past 20 years and
are being adapted by the paleoclimate community. See \citet
{tingley2010BARCAST1} for a specific example, and \citet
{tingley2010PiecingTogether} for a comprehensive discussion.

\section*{Acknowledgment}\label{sec:acknow}
This manuscript benefited from discussions with Peter Huybers.

\begin{supplement}%[id=suppA]
\stitle{Matlab code}
\slink[doi]{10.1214/10-AOAS398ESUPP} %[doi,text={...}] - jei reikia
%suskaldyti doi
\slink[url]{http://lib.stat.cmu.edu/aoas/398E/supplementE.zip}
\sdatatype{.zip}
\sdescription{A set of Matlab files that carry out the experiment
described in the text, and generate the figures.}
\end{supplement}

% imsref loaded by smiklovaite, 2011-01-21 09:06:47

\printaddresses

\end{document}